\documentclass{article}
\usepackage{dcase2024,amsmath,graphicx,url,times,booktabs, tabularx}
\usepackage{makecell}
\usepackage{array}
\usepackage{graphicx}
\usepackage{ragged2e}
\usepackage{xcolor}
\usepackage{hyperref}
\usepackage{cite}
\usepackage{siunitx}
\usepackage{multirow}
\usepackage{arydshln} 
\hypersetup{
    colorlinks=true,   
    urlcolor=blue,     
}

\title{A Reference-free Metric for Language-Queried Audio Source Separation using Contrastive Language-Audio Pretraining}

\name{Feiyang Xiao$^1$, Jian Guan$^{1*}$\thanks{*Corresponding author.}, Qiaoxi Zhu$^2$, Xubo Liu$^3$, Wenbo Wang$^4$, Shuhan Qi$^5$,
    \thanks{This work was partly supported by the project of the Ministry of Industry and Information Technology under Grant No.CBZ3N21-2.}
}
\secondlinename{Kejia Zhang$^1$, Jianyuan Sun$^6$, and Wenwu Wang$^3$}
\address{
    $^1$College of Computer Science and Technology, Harbin Engineering University, Harbin, China\\
    $^2$University of Technology Sydney, Ultimo, Australia\\
    $^3$Centre for Vision, Speech and Signal Processing, University of Surrey, Guildford, UK\\
    $^4$Faculty of Computing, Harbin Institute of Technology, Harbin, China\\
    $^5$School of Computer Science and Technology, Harbin Institute of Technology, Shenzhen, China\\
    $^6$Department of Computer Science, University of Sheffield, Sheffield, UK\\
}

\begin{document}

\ninept
\maketitle

\begin{sloppy}

\begin{abstract}

Language-queried audio source separation (LASS) aims to separate an audio source guided by a text query, with the signal-to-distortion ratio (SDR)-based metrics being commonly used to objectively measure the quality of the separated audio. However, the SDR-based metrics require a reference signal, which is often difficult to obtain in real-world scenarios. In addition, with the SDR-based metrics, the content information of the text query is not considered effectively in LASS. This paper introduces a reference-free evaluation metric using a contrastive language-audio pretraining (CLAP) module, termed CLAPScore, which measures the semantic similarity between the separated audio and the text query. Unlike SDR, the proposed CLAPScore metric evaluates the quality of the separated audio based on the content information of the text query, without needing a reference signal. Experiments show that the CLAPScore provides an effective evaluation of the semantic relevance of the separated audio to the text query, as compared to the SDR metric, offering an alternative for the performance evaluation of LASS systems. The code for evaluation is publicly available\footnote{GitHub: \url{https://github.com/LittleFlyingSheep/CLAPScore_for_LASS}}.

\end{abstract}


\section{Introduction}
\label{sec:intro}

Language-queried audio source separation (LASS) focuses on separating an audio source from a multi-source mixture based on a natural language description, i.e., a text query \cite{lass, audiosep}. Unlike traditional audio source separation, LASS utilizes the complex and rich semantic information of natural language to guide the separation process \cite{lass}. This integration of multi-modal data allows for more intuitive and flexible interaction with audio separation systems, making it particularly useful in various applications, i.e., audio editing \cite{wang2024consistent, liu2024audio, tan2023language, dong2023clipsep}, multimedia content creation \cite{liu2023audioldm}, and designs of assistive listening devices \cite{lass, audiosep, ResUNet, pons2024gass}.

Following audio source separation literature \cite{luo2018tasnet, xiao2021time, luo2019conv}, the signal-to-distortion ratio based metrics, i.e., SDR \cite{vincent2006performance}, SDR improvement (SDRi) \cite{scheibler2022sdr, chen2024mdx}, and scale-invariant SDR (SI-SDR) \cite{le2019sdr} have been used to measure the separation performance of LASS methods in \cite{lass}. All these metrics aim to quantify the quality of the separated audio signals. They measure how close the separated audio is to the original target audio, focusing on the reduction of distortion or errors introduced during the separation process \cite{scheibler2022sdr}. 

However, a major limitation of these SDR-based metrics is that they need a reference audio to compare against the separated audio. This makes these metrics applicable only in the simulated environments with known target audio, but impractical for real-world applications where the target source is unknown \cite{montresor2018reference}. In such cases, alternative evaluation methods or proxy measures are required to evaluate the performance of the audio separation algorithms.

\begin{figure*}[t]
    \centering
    \includegraphics[width=.98\textwidth]{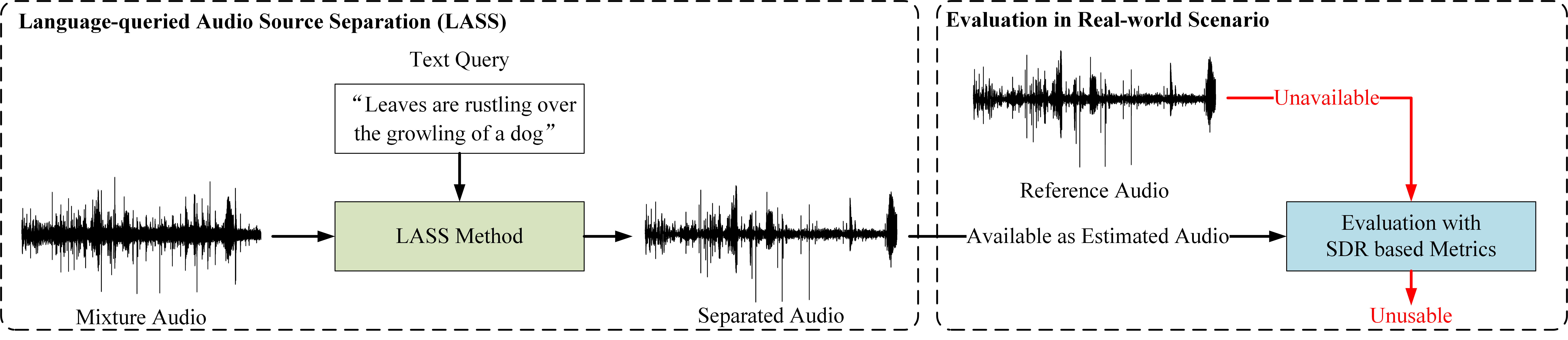}
    \vspace{-4mm}
    \caption{Illustration of the limitation of the SDR-based metrics for the evaluation of the language-queried audio source separation (LASS) methods in the real-world scenario, where the reference audio required by the SDR-based metrics is unavailable. Therefore, the SDR-based metrics are unusable for the evaluation of the LASS methods in the real-world scenario.}
    \label{fig:limiation_of_SDR}
\end{figure*}

In this paper, we introduce a reference-free evaluation metric for LASS, which calculates the audio-text similarity score using the contrastive language-audio pretraining (CLAP) module \cite{wu2023large}, termed CLAPScore. Unlike the previous SDR-based metrics that require a reference audio to measure the separation performance, the proposed CLAPScore metric evaluates the semantic similarity between the separated audio and the text query without needing a reference audio. This makes CLAPScore metric particularly useful for real-world applications where a reference audio may not be available. Furthermore, similar to SDRi, the improvement in CLAPScore (CLAPScore-i) from the mixture to the separated audio can reflect the improvement from LASS methods. Moreover, the CLAPScore is also expanded to incorporate the reference audio while it is available, denoted as RefCLAPScore.

Experiments indicate that the proposed CLAPScore metric exhibits an approximately linear correlation with the SDR metric, suggesting that CLAPScore can effectively evaluate the separation performance of the LASS methods. Additionally, since the CLAPScore metric does not require reference audio and relies solely on the text query used in the LASS separation process, it can be utilized to evaluate LASS in real-world scenarios where the reference audio is unavailable. This capability facilitates the development and evaluation of the LASS methods on real-world multi-source data.

\begin{figure*}[t]
    \centering
    \includegraphics[width=.98\textwidth]{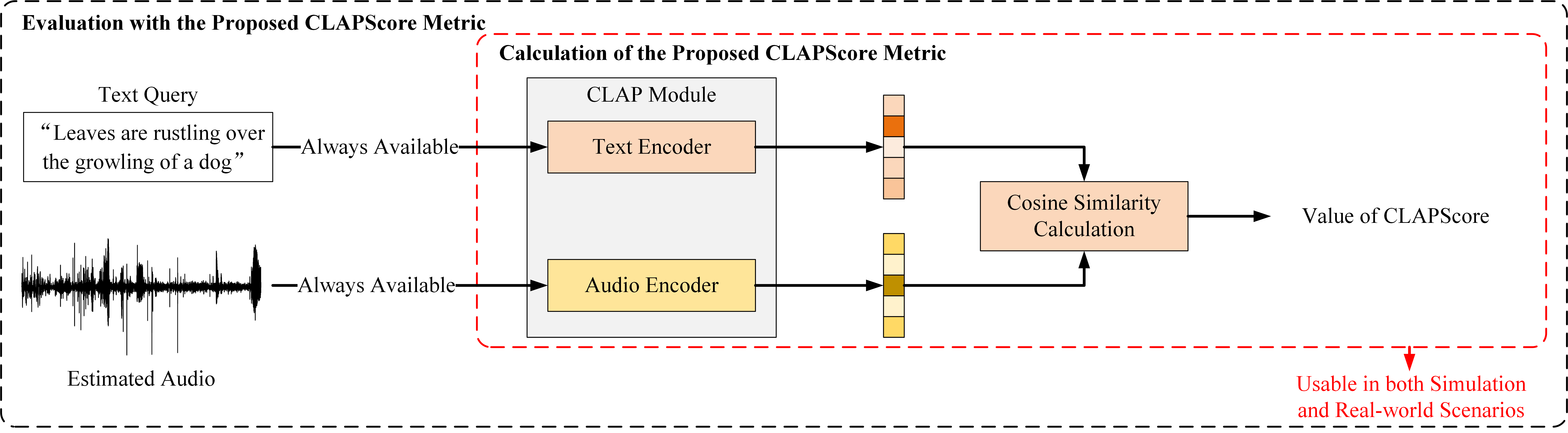}
    \vspace{-4mm}
    \caption{Illustration of the evaluation process with the proposed CLAPScore metric for language-queried audio source separation. Notably, the proposed CLAPScore metric does not need a reference audio for the evaluation. The inputs of the proposed CLAPScore metric, i.e., the estimated audio and the text query, are available in both simulation and real-world scenarios. Therefore, the CLAPScore metric can be applicable for both such scenarios.}
    \vspace{-4mm}
\label{fig:evaluation_with_ATS}
\end{figure*}

\section{Previous SDR-based Metrics}
\label{sec:related_work}

The SDR-based metrics (i.e., SDR, SDRi, and SI-SDR) are widely used objective metrics in signal processing, particularly in language-queried audio source separation \cite{scheibler2022sdr}. These metrics can provide a reliable and standardized method for evaluating the quality of the separated audio from LASS methods in the simulation scenario but are limited in the real world \cite{montresor2018reference}.

\subsection{Definition of SDR-Based Metrics}

In widely used SDR-based metrics, SDR measures the ratio of the power of the desired signal to the power of the distortion introduced by the separation process \cite{vincent2006performance}. SDRi is an improvement metric that measures the difference in SDR before and after applying an audio source separation algorithm \cite{scheibler2022sdr, chen2024mdx}. SI-SDR normalizes the audio signals to make the evaluation independent of their amplitude, which is more robust for varying scales \cite{le2019sdr, tzinis2020two, von2022sa}. The definition of SDR, SDRi and SI-SDR can be presented as follows:
\begin{equation}
\label{eq:sdr}
    \text{SDR} = 10 \log_{10} \left( \frac{\| \mathbf{s} \|^2}{\| \mathbf{s} - \hat{\mathbf{s}} \|^2} \right),
\end{equation}
\begin{equation}
\label{eq:sdri}
    \text{SDRi} = \text{SDR}_{\text{after}} - \text{SDR}_{\text{before}},
\end{equation}
\begin{equation}
\label{eq:sisdr}
    \text{SI-SDR} = 10 \log_{10} \left( \frac{\| \alpha \mathbf{s} \|^2}{\| \alpha \mathbf{s} - \hat{\mathbf{s}} \|^2} \right),
\end{equation}
where $\mathbf{s}$ denotes the reference audio, i.e., the ground-truth audio source, $\hat{\mathbf{s}}$ denotes the estimated audio. $\text{SDR}_{\text{before}}$ denotes the SDR between the mixture and the reference audio, and $\text{SDR}_{\text{after}}$ denotes the SDR between the separated audio from a LASS method and the reference audio. The improvement from $\text{SDR}_{\text{before}}$ to $\text{SDR}_{\text{after}}$ is the value of SDRi. For SI-SDR, $\alpha=\frac{\hat{\mathbf{s}}^{\top}\mathbf{s}}{\|\mathbf{s}\|^2}$ is the optimal scaling factor that aligns the estimated audio with the reference audio, where $\top$ denotes the transpose operation. For all of these SDR-based metrics, a higher value indicates better separation performance. 

\subsection{Limitation of SDR-Based Metrics}

According to the above definition of SDR-based metrics, it can be found that, these metrics all depend on the reference audio signal $\mathbf{s}$ to measure the separation performance of the LASS methods. However, this requirement can be only met in a simulation scenario, where the reference audio and the noise are known to simulate the mixture audio. Due to the lack of the reference audio, these SDR-based metrics cannot be usable to measure the LASS performance in the real-world scenario, as illustrated in Figure~\ref{fig:limiation_of_SDR}. 

Moreover, these SDR-based metrics are power-based metrics to measure the effectiveness of LASS methods. They primarily focus on the signal quality and distortion level of the separated audio, without considering whether the semantic content of the separated audio matches the text query. Therefore, these SDR-based metrics cannot measure the semantic similarity between the separated audio and the text query. To measure the matching of the semantic content between the separated audio and the text query, other more effective semantic similarity metrics are required. 

\section{Proposed CLAPScore Metric}

To measure how well the separated audio matches the text queries, we introduce the CLAPScore metric. This metric quantifies how closely the content of the separated audio aligns with the text query. A higher CLAPScore means that the separated audio's content is more similar to the text query, indicating better performance in separating audio based on the text query. The evaluation process with the proposed CLAPScore metric is illustrated in Figure~\ref{fig:evaluation_with_ATS}. 

\subsection{Definition of Proposed CLAPScore Metric}

The proposed CLAPScore metric is a measure of the similarity between the separated audio from the LASS methods and the text query used in the LASS process. It can measure the semantic similarity between the separated audio and the text query.

The calculation of the proposed CLAPScore metric is based on the contrastive language-audio pretraining (CLAP) module \cite{wu2023large}. The CLAP module is pretrained on a large-scale dataset and learns the audio-text alignment in the latent space \cite{wu2023large}. Due to this advantage, the CLAP module is widely used to measure the audio-text alignment in the evaluation of text-to-audio generation methods \cite{huang2023make, liu2024audioldm}. Inspired by these studies, we introduce the CLAP module to calculate the audio-text similarity between the estimated audio and the text query to measure the separation performance of the LASS methods.

Specifically, the audio embedding of the estimated audio $\hat{\mathbf{s}}$ (i.e., the separated audio signal) and the text embedding of the text query are obtained with the CLAP module\footnote{\url{https://huggingface.co/spaces/Audio-AGI/AudioSep/blob/main/checkpoint/music_speech_audioset_epoch_15_esc_89.98.pt}}, as follows,
\begin{equation}
    \hat{\mathbf{a}} = E_{A} (\hat{\mathbf{s}}),
\end{equation}
\begin{equation}
    \mathbf{t} = E_{T} (\mathbf{c}),
\end{equation}
where $\mathbf{c}$ denotes the text query, $E_{A} (\cdot)$ and $E_{T} (\cdot)$ denotes the audio encoder and text encoder in CLAP module, respectively. The audio embedding $\hat{\mathbf{a}}$ of the estimated audio is extracted by the audio encoder in the CLAP module, and the text embedding $\mathbf{t}$ of the text query is extracted by the text encoder in the CLAP module.

Then, the cosine similarity between the audio embedding and the text embedding is calculated as the value of the proposed CLAPScore metric to measure the semantic similarity between the estimated audio and the text query. Thus, the calculation of the audio-text similarity score can be represented as
\begin{equation}
\label{eq:ats}
    \text{CLAPScore} = \frac{\hat{\mathbf{a}}^{\top} \mathbf{t}}{\| \hat{\mathbf{a}} \| \| \mathbf{t}\|}.
\end{equation}
A higher CLAPScore means a better match between the audio embedding of the estimated audio and the text query used in LASS process. Therefore, a higher CLAPScore indicates better separation performance of the LASS methods. 

\subsection{Advantages of the Proposed CLAPScore Metric}

Different from the SDR-based metrics, the proposed CLAPScore metric can evaluate the degree of matching between the separated audio and the text query in their latent spaces. It provide a way to measure the semantic similarity between the separated audio and the text query for the LASS task.

In addition, according to the definition of the proposed CLAPScore metric, it can be found that, the evaluation based on the proposed CLAPScore metric depends on the separated audio and the text query, without the need for a reference audio as required in the SDR-based metrics. The separated audio and the text query can be easily obtained in both the simulation and the real-world scenarios, thus this metric is applicable for both scenarios, offering advantages over the SDR-based metrics which only work when the reference audio is available.

\subsection{Expanded CLAPScore Improvement Metric}

In addition, similar to the SDRi metric, we design the improvement of the CLAPScore metric to measure the difference in the proposed CLAPScore metric before and after applying an LASS method, termed CLAPScore improvement (CLAPScore-i). The CLAPScore-i metric can be calculated as follows,
\begin{equation}
    \text{CLAPScore-i} = \text{CLAPScore}_{\text{after}} - \text{CLAPScore}_{\text{before}},
\end{equation}
where $\text{CLAPScore}_{\text{before}}$ denotes the CLAPScore between the original mixture audio and the text query, and $\text{CLAPScore}_{\text{after}}$ denotes the CLAPScore between the separated audio and the text query. 

\subsection{Expanded RefCLAPScore Metric}

We present an expanded CLAPScore while the reference audio is available, termed RefCLAPScore. The calculation of the RefCLAPScore can be represented as 
\begin{equation}
    \text{RefCLAPScore} = H(\text{CLAPScore}_{\text{after}}, \text{CLAPScore}_{\text{ref}}),
\end{equation}
where $H(\cdot, \cdot)$ denotes the harmonic mean function, and $\text{CLAPScore}_{\text{ref}}$ denotes the CLAPScore of the reference audio. The RefCLAPScore metric can further introduce the semantic information of the reference audio (i.e., source audio) to obtain a fine-grained measure for the separation performance.

\section{Experiments}

\subsection{Dataset}

To verify the effectiveness of the proposed CLAPScore metric, we conducted experiments on the DCASE 2024 Challenge Task 9 validation set\footnote{\url{https://zenodo.org/records/10886481}}. This dataset includes 1000 audio signals from the FreeSound dataset \cite{fonseca2017freesound}, each with 3 corresponding text queries. By randomly combining pairs of audio signals, the validation set provides 3000 mixture audio samples for evaluation. Additionally, we split this dataset to perform an ablation study of the proposed CLAPScore metric.

\subsection{Effectiveness of Proposed CLAPScore Metric}

To demonstrate the effectiveness of the proposed CLAPScore metric, we evaluate the separation performance of standard LASS methods on 3000 officially provided mixture audio signals using both SDR-based metrics (SDR, SDRi, SI-SDR) and CLAPScore based metrics (CLAPScore, CLAPScore-i, RefCLAPScore). The evaluated LASS methods include the official baseline of the DCASE 2024 Challenge Task 9 (baseline) \cite{audiosep}, our previously submitted system \cite{guan2024_t9} trained with GPT-augmented text queries (baseline-Augmented) \cite{primus2023_t6b, Primus2023}, and the state-of-the-art method, AudioSep \cite{audiosep}. Evaluation results measured by these metrics are shown in Table~\ref{tab:LASS_performance}.

Based on the SDR metric performance, it is clear that the separation effectiveness of the three evaluated LASS methods ranks from highest to lowest as follows: AudioSep, baseline-Augmented, and baseline. Similarly, in the evaluation using the CLAPScore metric, the methods rank from best to worst in the same order: AudioSep, baseline-Augmented, and baseline. This demonstrates that the CLAPScore metric can effectively assess the separation performance of LASS methods. Furthermore, its ability to evaluate without requiring a reference audio makes it particularly suitable for scenarios where reference audio is unavailable.

\begin{table}[t]
    \centering
    \caption{Evaluation of different LASS methods with the SDR-based metrics (i.e., SDR, SDRi, SI-SDR) and the proposed CLAPScore based metrics (i.e., CLAPScore, CLAPScore-i, RefCLAPScore).}
    \vspace{0.5em}
    \resizebox{\linewidth}{!}{
    \begin{tabular}{ccccccc}
        \toprule
        Method & SDR & SDRi & SI-SDR & CLAPScore & CLAPScore-i & RefCLAPScore \\
        \midrule
        Baseline \cite{audiosep} & 5.708 & 5.673 & 3.862 & 0.239 & 0.029 & 0.253 \\
        Baseline-Augmented \cite{guan2024_t9} & 5.937 & 5.902 & 4.191 & 0.242 & 0.031 & 0.254 \\ 
        AudioSep \cite{audiosep} & \textbf{8.192} & \textbf{8.157} & \textbf{6.680} & \textbf{0.261} & \textbf{0.050} & \textbf{0.267} \\
        \bottomrule
    \end{tabular}
    }
    \vspace{-4mm}
\label{tab:LASS_performance}
\end{table}

\subsection{Correlation between SDR-Based Metrics and CLAPScore}

According to the results in Table~\ref{tab:LASS_performance}, an interesting phenomenon can be observed that the performance measured by CLAPScore based metrics (i.e., CLAPScore, CLAPScore-i, and RefCLAPScore) shows similar trend to that measured by SDR-based metrics. Specifically, when the performance on CLAPScore based metrics is high, the performance on SDR-based metrics is also high. To explore their correlation, we calculate the Pearson correlation coefficient (PCC) as Table~\ref{tab:correlation}.

It can be found that, both CLAPScore and RefCLAPScore shows a moderate positive correlation with both SDR and SI-SDR. Additionally, CLAPScore-i has a similar moderate correlation with SDRi. These indicate that the CLAPScore based metrics has statistically significant positive correlations with SDR-based metrics.

To further explore the correlation between these metrics, we simulate the mixture audio under different SDR levels ranging from $-$20dB to 20dB in 5dB increments, based on the provided 3000 source-noise pairs in the validation set of DCASE 2024 Challenge Task 9. Then, we evaluate the quality of these simulated mixture audio and the quality of the separated audio from the LASS method (i.e., AudioSep \cite{audiosep}) using the proposed CLAPScore based metrics. The results are illustrated in Figure~\ref{fig:relation_between_SDR_and_ATS}.

\begin{table}[t]
    \centering
\caption{Pearson correlation coefficient (PCC) between SDR-based and CLAPScore-based metrics with statistically significant correlation p-value $< 0.05$.}
\vspace{0.5em}
    \scriptsize
    \resizebox{\linewidth}{!}{
    \begin{tabular}{ccccc}
    \toprule
                     & PCC with SDR      & PCC with SI-SDR   &                              & PCC with SDRi \\
       \midrule
        CLAPScore    & 0.270 & 0.289 & \multirow{2}{*}{CLAPScore-i} & \multirow{2}{*}{0.288} \\
        RefCLAPScore & 0.226 & 0.254 & &  \\
        \bottomrule
    \end{tabular}
    }
    \vspace{-4mm}
\label{tab:correlation}
\end{table}

\begin{figure}[t]
    \centering
    \includegraphics[width=1.0\linewidth]{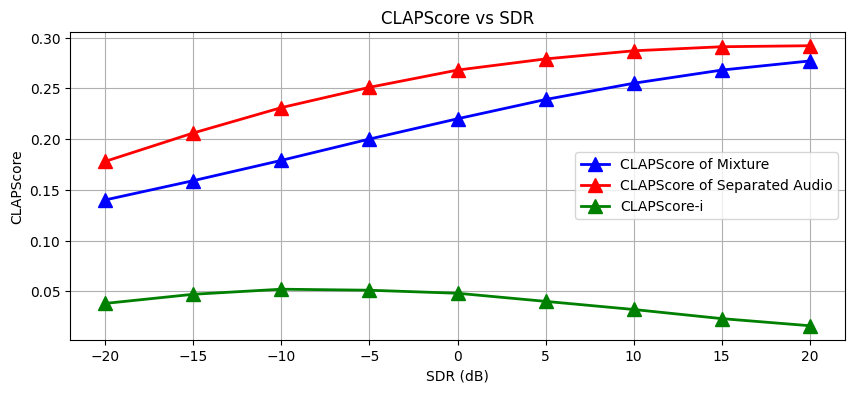}
    \vspace{-7mm}
    \caption{Illustration to show the correlation between the SDR metric and the proposed CLAPScore metric. Here, the separated audio comes from the LASS method, i.e., AudioSep \cite{audiosep}.}
    \vspace{-2mm}
\label{fig:relation_between_SDR_and_ATS}
\end{figure}

The proposed CLAPScore for mixture audio shows an approximately linear correlation with the SDR metric, as shown by the blue line in Figure~\ref{fig:relation_between_SDR_and_ATS}. This indicates that CLAPScore effectively evaluates audio signal quality using text queries. Additionally, Figure~\ref{fig:relation_between_SDR_and_ATS} demonstrates that the CLAPScore for separated audio (red line) and CLAPScore-i (green line) indicate a better match with text queries for separated audio, validating CLAPScore's effectiveness in measuring separated audio quality. Notably, CLAPScore-i for AudioSep is higher at lower SDR levels than at higher SDR levels, likely because simulated mixtures at higher SDR levels are already close to the source audio, resulting in only subtle improvements with the LASS method.

\subsection{Evaluation with Different Mixing Strategies}

We conduct an ablation study to evaluate the CLAPScore value of the mixture audio signals with different mixing strategies, where 990 audio signals are selected from the validation set of DCASE 2024 Challenge Task 9 as source audio and three different mixing strategies are attempted for each source audio: (1) source audio, (2) mixed with white noise, and (3) mixed with an audio signal of different content. This results in a total of 2970 mixtures for evaluation, with each mixing strategy producing 990 estimated audio signals. The lines representing the CLAPScore metric at different SDR levels ($-$20dB, $-$15dB, $-$10dB, $-$5dB, 0dB, 5dB, 10dB, 15dB, and 20dB) for these mixtures are shown in Figure~\ref{fig:ATS_with_different_noises}.

\begin{figure}[t]
    \centering
    \includegraphics[width=1.0\linewidth]{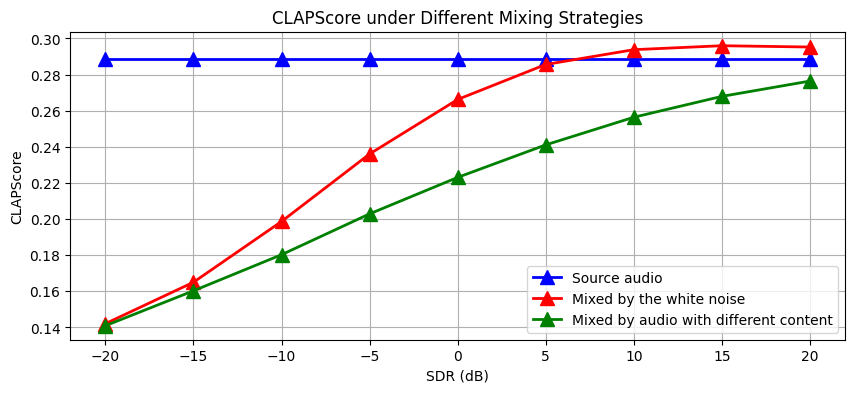}
    \vspace{-7mm}
    \caption{Illustration of the proposed CLAPScore metric for the mixtures from different mixing strategies.}
    \vspace{-4mm}
\label{fig:ATS_with_different_noises}
\end{figure}

It can be found that, the value of the proposed CLAPScore for the source audio is significantly better than the one mixed by audio with different content, under any SDR levels. This verifies that the proposed CLAPScore metric can capture the difference on the semantic content between the estimated audio and the text query. Therefore, the proposed CLAPScore metric prefers to assign an estimated audio that has different content from the text query with a lower measure, even if the SDR performance of the estimated audio is good (i.e., 20dB).

Furthermore, it is interesting that the estimated audio mixed with the white noise has higher CLAPScore value than the original source audio under high SDR levels (i.e., 10dB, 15dB, 20dB). The reason may be that, in these SDR levels, the white noise can be considered as the background noise, estimated audio mixed by such background noise may enhance the realism of the resulting mixes, as analyzed in \cite{pons2024gass}. Then, the enhanced realism of the estimated audio leads to better CLAPScore performance than the source audio.

\section{Conclusion}

In this work, we proposed a reference-free metric for language-queried audio source separation using contrastive language-audio pretraining, termed CLAPScore, which can further measure the semantic similarity between the estimated audio and the text query, without the requirement of a reference audio. Experiments show that the proposed CLAPScore can achieve a more fine-grained evaluation for language-queried audio source separation.  

\bibliographystyle{IEEEtran}
\bibliography{refs}

\end{sloppy}
\end{document}